\documentclass[11pt]{article}
\usepackage{techart}
\pdfoutput=1

\renewcommand\vec[1]{\ensuremath{\boldsymbol{#1}}}

\newcommand\xa{\ensuremath{\vec{x}_a}}
\newcommand\xb{\ensuremath{\vec{x}_b}}

\newcommand\za{\ensuremath{\vec{z}_a}}
\newcommand\zb{\ensuremath{\vec{z}_b}}

\begin{document}
\title{On a recent development in stochastic inversion\\
with applications to hydrogeology}
\date{}
\author{Zepu Zhang}
\maketitle

\barefoot{%
Copyright \copyright\@ 2011, 2012 Zepu Zhang. All rights reserved.}

\vspace*{-5mm}

\begin{abstract}
We comment on a recent approach to spatial stochastic inversion,
which centers on a concept known as ``anchors''
and conducts nonparametric estimation of the likelihood of the anchors
(along with other model parameters) with respect to data obtained
from field processes.
Conceptual and technical observations are made
regarding the development, interpretation, and use of this approach.
We also point out that this approach has been superseded by new developments.
\medskip

\end{abstract}
\bigskip

\citet{Zhang:2008:DAS} introduced
a new stochastic approach, named ``anchored inversion'',
to a type of spatial inverse problems.
Conceptually,
the approach is centered on a parameterization device
that ``transforms information'' contained in ``indirect data'' to the
form of ``direct data''.
Technically and computationally,
it is centered on a nonparametric procedure for estimating
multivariate likelihood.
The approach was fully described and implemented in
\citet{Zhang:2008:IMS1},
which also saw adoption of the name
``anchor'' for the parameterization device
and ``anchored inversion'' for the methodology.

This method was re-presented in \citet{Rubin:2010:BAI}
by massive copy-pasting and rephrasing
without the knowledge, consent, and acknowledgement of Zhang
(see \url{http://www.AnchoredInversion.info}).
The re-presentation contained numerous technical errors and conceptual
misinterpretations.
In addition, key steps of the technical core,
namely nonparametric estimation of likelihood,
were omitted, rendering the description severely incomplete.

In this note,
we point out some of the errors in \citet{Rubin:2010:BAI},
clarify concepts in the method,
fill in the missing technical component,
and make additional comments on the method.
It is hoped that this effort will help reduce
misunderstandings about the method caused by the
technically flawed presentation of \citet{Rubin:2010:BAI}.

In the end we point out that the method described in
\citet{Zhang:2008:IMS1} (or \citet{Rubin:2010:BAI})
has been superseded by a radically different, and superior, method.

\section{Concepts of data classification and anchors}

The design philosophy of the method is explained in the
sections 5 and~9 of \citet{Zhang:2008:IMS1};
the technical details of the method are given in the
sections 3, 4, 6 and the Appendix of \citet{Zhang:2008:IMS1}.
Also see \citet{Zhang:2008:MNM}
(which uses the less preferred name ``method of anchored
distributions'').
These materials are not to be repeated here.

Following the notation of \cite{Zhang:2008:IMS1},
$Y$ refers to the spatial attribute of interest
(such as hydraulic conductivity), $x$ indicates spatial location.
Geostatistical (or ``structural'') parameters are denoted by
$\theta$; anchors are denoted by $\vartheta$.
The forward process (or model) is denoted by $\mathcal{M}$.

\citet{Zhang:2008:IMS1} classifies on-site data into two types.
Type~A data ($\za$) are direct values of $Y$ at individual locations.
Type~B data ($\zb$) are functions of $Y$ at multiple locations (usually the
entire domain).
Type~A data are so defined that they can be used directly
in creating conditional fields
(see (A-3) in \citet{Zhang:2008:IMS1},
or (13) in~\citet{Rubin:2010:BAI}).
As for type~B data,
it is important to note that they are functions of the $Y$ field.
One can not establish correspondence between a type~B data component
and the $Y$ values (or anchors) at specific locations.
Moreover,
type~B data may not be ``spatial'' or possess measurement ``locations''
in the usual sense.
For example,
the data may be a time series of tracer concentrations in a transport
experiment, measured at one or more locations;
or they may even be a summary of such a time series.
Measurement locations, if any, of type~B data
are an internal business of the forward model $\mathcal{M}$.
They are not a concern to the inversion procedure.
What the inversion procedure expects is a prediction vector
$\mathcal{M}(\tilde{y})$,
produced by the forward model for any field realization $\tilde{y}$,
that can be compared with the observed data vector,
$\zb$.

In the definition of type~A data,
\citet{Rubin:2010:BAI} calls $y$ a ``known function''
(below formula~(1)).
This is inaccurate.
Here,
the data must be the spatial attribute $Y$ itself
(at locations $\xa$) and not any transform
thereof, because otherwise the use of the data $\za$ in all subsequent
formulas needs to be modified.
However, the data may have been obtained through transformations.
For example,
one may arrive at an estimate of the hydraulic conductivity
at location $x$ using empirical relations that connect conductivity with
core-support geophysical properties obtained at that location.
In this case, the type~A data $\za$ are the estimated conductivity,
rather than the geophysical property (which is an empirical function of
conductivity).
The transformation may well be nonlinear.

At times \citet{Rubin:2010:BAI} is confused about the concept of
type~B data and talks about anchors ``corresponding to'' specific
type~B data components; see, e.g., paragraphs [18], [58], [82].
A further inaccuracy appears in paragraph~[44]
of \citet{Rubin:2010:BAI}, which states that
``$M(\tilde{y})$ could be used to generate the field or possibly a
number of fields $\tilde{z}_b$''.
This statement is misleading in that $\tilde{z}_b$ (value
of the forward model) is \emph{not} a ``field''.
The use of the symbol $\tilde{z}_b$ here
(likewise in paragraph~[36], and in the ``Notation'' with entry
``$\tilde{z}_b$'')
is confusing,
noticing that it has been declared in
paragraph~[15] that ``the tilde sign over a variable denotes a field of
that variable''.
It is noted that the type~B data (i.e.\@ $\zb$) need not be anything
like a ``field'' in the meaning that is familiar to the audience of that article.
For example, the vector $\zb$ may be the combination of a few head
measurements and a time series dataset of tracer concentration.
Besides,
$\mathcal{M}(\tilde{y})$, given a $\tilde{y}$,
can generate only one $\tilde{z}_b$, not ``a number of'' $\tilde{z}_b$.
The reason is that the forward function is deterministic with
$\tilde{y}$.


Overall,
the role of anchors is to connect data (both types A and B)
with the unknown field $Y$.
\citet[paragraph~{[23]}]{Rubin:2010:BAI} mistakenly mentions
the role of the anchors as a
``mechanism for connecting between type~A and type~B data''.

A conceptual justification for the use of anchor
is given in \citet[sec.~3.3]{Zhang:2008:DAS}.
The idea is that (after appropriate choice of anchors and adequate
inference of their posterior distribution) anchors ``capture''
the information contained in data $\zb$ about the field $Y$,
so that simulation of $Y$ conditional on $z_b$
can be substituted by simulation conditional on the anchors.
\citet[first paragraph of sec.~5]{Zhang:2008:IMS1} explains this idea as
follows.
\begin{quote}
A common goal in inverse modeling is to create simulations of the field
$\tilde{Y}$ that are ``conditioned'' on available data...
When the available data include type-B data, a common practice...\@ generates
conditional simulations...\@by sampling the conditional
distribution of the field, $p(\tilde{y} \given \theta, \vartheta)$...
There are conceptual difficulties in this treatment, to be discussed in
section~8.4.
A natural reason for this treatment is the absence of
an intermediate device.
This device would
guarantee that fields simulated with the help of this device
are already conditioned on the type-B data,
therefore there is no need to verify (or reject)
the conditioning by evaluating the function $\mathcal{M}$ on the created fields.
\end{quote}
This explanation is re-used in \citet[paragraph~{[16]}]{Rubin:2010:BAI}.
More details along this line are given in
\citet[secs.~5,~6.3]{Zhang:2008:IMS1}, which are reused in
\citet[secs.~2.2,~5]{Rubin:2010:BAI}.
However, there are issues with this justification.


\section{Numerical estimation of likelihood}

The computational core of the method
is numerically estimating the likelihood
$p(z_b \given \theta, \vartheta)$.
The central position of this likelihood term
in the entire methodology is evident in
formula~(6) and Figure~1 of \citet{Zhang:2008:IMS1}
(or formula~(3) and Figure~1 of \citet{Rubin:2010:BAI}).
That this likelihood is nonparametrically, numerically estimated
is a defining feature of anchored inversion.
The other components such as prior specification, simulation,
and prediction are not unique to this method.

The importance of the numerical estimation of this likelihood is also
shown in \citet{Rubin:2010:BAI} at multiple places,
such as in paragraphs
[23], [33]--[37], [73], [121].
However,
\citet{Rubin:2010:BAI} does not provide technical details
of this component.
Even the dedicated section~3.2 only contains
a sketchy verbal description in paragraph~[36],
which is borrowed from
\citet[paragraph~4 of sec.~6.2]{Zhang:2008:IMS1}.
(The same description also appears in
\citet[paragraph~5 of sec.~3.1.1]{Murakami:2010:BAT}
and
\citet[sec.~2.3.4]{Murakami:2010:DBG}).
This unfortunate omission creates an awkward situation:
the article by \citet{Rubin:2010:BAI}
presents a methodology without presenting how it works.

The numerical procedure of likelihood estimation
adopted in \citet[sec.~7.2]{Zhang:2008:DAS} and
\citet[sec.~6.2]{Zhang:2008:IMS1}
is the $k$-th nearest neighbor estimator.
Let
$\vec{z}_1$,..., $\vec{z}_n$ be an i.i.d.\@ sample from the
$d$-variate (continuous) distribution of $\vec{Z}$,
whose density function is denoted by $f$.
The density at $\zb$ (i.e.\@ the observed type~B data vector) is
\begin{equation}\label{eq:density-multivar-hist}
f(\zb)
= \lim_{r \to 0}
    \frac{\sum_{i=1}^n I(\lvert \vec{z}_i - \zb \rvert \le r)}
         {n}
    \frac{1}{v_d\, r^d}
,\quad
\text{as }
n \to \infty
,
\end{equation}
where $I(\cdot)$ is the identity function,
$|\cdot|$ is the Euclidean distance, and
$v_d = \pi^{d/2} / \Gamma(1 + d/2)$ is the volume
of the unit hypersphere of dimension $d$.
Denote by $r$ the distance between $\zb$
and its $k$-th nearest neighbor in the sample $\{\vec{z}_i\}_1^n$.
Then an estimate of the density is
\begin{equation}\label{eq:density-multivar-knn}
\hat{f}(\zb) = \frac{k - 1}{n}\, \frac{1}{v_d\, r^d}
.
\end{equation}
As $n \to \infty$ and $r \to 0$,
this estimate approaches the true density.
\citet{Loftsgaarden:1965:NEM} suggests $\sqrt{n}$
as an empirical choice of $k$.

There is a large statistical literature on multivariate density estimation.
(``Likelihood'' is a probability density.)
It is not difficult to adjust specifics of this procedure,
but the spirit of nonparametric, numerical estimation remains.

\citet[paragraphs~{[37]},~{[109]}]{Rubin:2010:BAI} points out,
using the observations in
\cite[list after paragraph 4 in sec.~6.2]{Zhang:2008:IMS1},
that only the likelihood at the observed $z_b$ value
needs to be estimated;
moreover, the estimation may use a large sample size $n$
as long as computational resources permit.
\citet[paragraph~{[37]}]{Rubin:2010:BAI} also cites the relatively low
dimensionality of $(\theta, \vartheta)$ as a factor that contributes to
the method's computational efficiency compared with ``full-grid
inversion''.
However, anchored inversion and ``full-grid inversion'' use their
respective parameter vectors,
$(\theta, \vartheta)$ versus $\tilde{y}$,
in different ways; their relative computational cost
is unrelated to the comparison of their parameter dimensionalities.

As mentioned above,
it is a signature of the method that
no parametric form is assumed for the likelihood function
$p(z_b \given \theta, \vartheta)$;
rather, the likelihood is estimated nonparametrically.
This decision is necessitated by the largely intractable model errors;
see \citet[sec.~8.4]{Zhang:2008:IMS1} for an explanation.
\citet[paragraphs {[23], [34--35]}]{Rubin:2010:BAI} states that
the likelihood $p(z_b \given \theta, \vartheta)$ may be taken to be
either parametric or nonparametric.
The application paper by
\citet{Murakami:2010:BAT}
actually makes an attempt to assume Gaussian likelihoods.
However,
it is important to note that the distribution of the forward process
output is dependent on
(1) the nature of the particular process;
(2) the field parameter ($\theta$ and $\vartheta$).
The distribution (both shape and parameters)
may change with any change of either of these two factors.
Besides,
it is a misconception that parametric methods are more efficient than 
nonparametric ones.
This is the case only when the parametric form is correct.

On the other hand,
if one does assume a parametric form for the likelihood function
(with known parameter values, whether or not the assumptions are
reasonable),
then the problem falls squarely in the realm of Markov chain Monte Carlo
(MCMC) and related sampling schemes.
In that case there is no advantage in using the newly proposed method.

\section{Specification of prior}

The anchored inversion methodology is open and flexible with respect to
the geostatistical formulation and the prior specification for the model parameters,
$\theta$ and $\vartheta$.
Nevertheless,
these technical details \emph{are} important because
(1) one has to make choices for these technicalities in order to
get the procedure running, and (2) the details affect the performance.
Of these technicalities, the specification of prior is the most important.

\citet{Zhang:2008:DAS} and \citet{Zhang:2008:IMS1}
adopt a geostatistical formulation that includes parameters
for trend, variance, scale, nugget, and the Mat{\'e}rn correlation.
In addition, Box-Cox transform is performed to deal with possible
nonnormality of $Y$.
The prior for the geostatistical parameters $\theta$
is given by formula~(A-2) in \citet{Zhang:2008:IMS1}.
This prior has been proposed by \citet{Pericchi:1981:BAT} to work with
the Box-Cox transform.
This prior is re-used by
\citet[formula~(11)]{Rubin:2010:BAI}.
However, \citet{Rubin:2010:BAI} choses to drop the Box-Cox transform.
As a result, the reason for using this particular form of prior is lost.
The same problem exists in
\citet[sec.~2.5.1]{Murakami:2010:DBG}.

\section{The Mat{\'e}rn covariance model}

The Mat{\'e}rn covariance (or correlation) model
is an important tool in spatial statistics.
\citet[sec.~3.2]{Zhang:2008:DAS} and
\citet[Appendix]{Zhang:2008:IMS1} include brief introductions to this tool.
An especially valuable asset of the Mat{\'e}rn model
is its parameter $\kappa$,
which concerns the smoothness of the
random field bearing this spatial association modeled by the Mat{\'e}rn
covariance function.
As $\kappa$ increases, the field becomes smoother in a certain
statistical sense.
One can choose a particular value of $\kappa$ based on this
interpretation, as is done in \citet{Zhang:2008:IMS1}.
Examining the full range of $\kappa$ is usually not a good idea:
it is both unnecessary and infeasible.
It is infeasible because it necessitates numerical evaluation of the
Bessel function (admittedly this is not a huge obstacle in this age).
It is unnecessary because one or a few reasonable smoothness levels
($\kappa$ values) usually provide adequate characterization
of ones random field under consideration (as far as smoothness is
concerned).
A deeper reason for skipping a ``full-range'' examination
lies in the intimate interactions between
the smoothness, scale (or range), and variance parameters;
the interactions cause difficulties in identifiability.

What is debatable in the use of the Mat{\'e}rn model
in \citet{Nowak:2010:BGD} and \citet{Rubin:2010:BAI}
is that both articles advocate the notion that
the full range of $(0, \infty)$ be explored for $\kappa$,
and this act be interpreted as ``model averaging''.
The authors emphasize the notion that
different values of $\kappa$ can be viewed as different covariance
models.

If one limits $\kappa$ to several discrete values, then the posterior of
$p(\kappa)$
places certain weight
on each of these values, hence the estimated correlation structure can
be viewed as a weighted
average of these special cases. If one picks a single estimate, it
amounts to a selection from
these several special cases. This model averaging or selection is
accomplished via parameter
estimation (of $\kappa$), as opposed to some external measure that
distinguishes the performance
of those special cases considered. Such model averaging interpretation
is less useful when $\kappa$ is
treated as a continuous variable, because the continuously changing
$\kappa$
represents a coherent
concept of ``smoothness'', and thinking of each arbitrary value of
$\kappa$ as a
special ``model''
in its own right is not beneficial. Another situation where the ``model
averaging/selection''
perspective is less useful is when the correlation function is not the
sole focus of the discussion
but rather constitutes part of a large parameter vector.

In the effort to introduce the Mat{\'e}rn model and promote the
``model selection/averaging'' interpretation,
\citet[sec.~4]{Rubin:2010:BAI} makes some statements that harm
the coherency of the description of the inversion method.
In paragraph~[47], for example, the article
confuses the ``forward model'' with the ``covariance model''.
It refers to any particular value of the parameter vector
$(\theta, \vartheta)$ as a particular (forward) ``model'',
and talks about the ``probability'' of each such model,
while $(\theta, \vartheta)$ in MAD is the model parameter and as such
certainly can take different values according to its posterior
distribution.
In paragraph~[48] it is claimed that
``equation~(3)... could be written for each of the $N$
combinations of $\theta_i$, $\vartheta_i$
instead of a single $\theta$, $\vartheta$ combination''.
However, equation~(3) is about \emph{parameters}
$\theta$ and $\vartheta$,
which can take various values (or ``combinations'') in the first place.
The equation is never about ``a single $\theta$, $\vartheta$ combination''.
For a single $\theta$, $\vartheta$ combination,
one needs only to plug the specific values into the equation.

\section{Choice of anchor locations}

Since anchors are parameters defined by the user,
the user needs to make decisions on how many anchors to use and where to
place them.

\citet[sec.~7.1]{Zhang:2008:DAS}
outlines factors that should be considered while determining the
placement of anchors.
Further details are given in
\citet[sec.~6.3 and paragraphs~6--7 of sec.~9]{Zhang:2008:IMS1}.
To summarize the main points,
(1) anchors should be placed where they are sensitive to the forward
process $\mathcal{M}$ so that information in the forward data
$z_b$ is efficiently ``transferred'' to anchors;
(2) anchors should be placed where one's future prediction tasks
need them most.
The second point is referred to as ``objective oriented inversion''
in \citet[paragraph 6 of sec.~9]{Zhang:2008:IMS1}.
Of these two considerations, the second is more important
because one's goal is never to ``re-play'' the
data-generating forward process $\mathcal{M}$,
whose outcome is already known.
When future prediction task is not yet known,
one should aim for the anchors to achieve overall
``good'' and balanced representation of the entire field $Y$,
with attention to regions possessing critical features.

These considerations also appear in
\citet[sec.~5]{Rubin:2010:BAI}
(in paragraphs [58] and~[62] the authors re-term the second criterion
``targeted anchor placement''),
\citet[sec.~3.1.4]{Murakami:2010:BAT},
and
\citet[sec.~3.4.4]{Murakami:2010:DBG}.

Regarding the ``right'' or ``optimum'' number of anchors to use,
Zhang [personal correspondence to Murakami and Rubin, 4 October 2008] suggests,
``The number of anchors should be determined based on a measure 
compromising the `sharpness' of conditioning and the identifiability of 
the parameters (anchor values). With a small number of anchors, the 
conditioning can't be very sharp, because the variability in the field 
conditioning [conditional] on the anchors is still large.''
Zhang [personal correspondence to Rubin, 21 December 2008]
further clarifies,
``Basically in the area of interest, we can
distribute inverted anchors according to some recommended sampling
design. Then we can increase the number of inverted anchors and get
prediction results. The right number of anchors is achieved when the
prediction stabilizes.''
These suggestions are rephrased in
\citet[paragraphs~{[59--60]}]{Rubin:2010:BAI}
and
\citet[sec.~2.3.5]{Murakami:2010:DBG}
without implementation.
Implementation of this idea is likely to be nontrivial.

\section{Sequential assimilation of multiple datasets}
\label{sec:updating}

\citet[sec.~9]{Zhang:2008:DAS} mentions the possibility of
incorporating multiple type~B datasets in a sequential manner,
and outlines the conditions under which this is or is not suitable.
This idea also appears in
\citet[sec.~7]{Rubin:2010:BAI}.
The formula~(16) in \citet{Rubin:2010:BAI}
is correct up to the second line.
The third ``$=$'' does not hold because
\[
p\Bigl(z_b^{(2)} \Bigm\lvert\, z_b^{(1)}, \theta, \vartheta_a,
    \vartheta_b^{(1)}, \vartheta_b^{(2)}\Bigr)
\ne
p\Bigl(z_b^{(2)} \Bigm\lvert\, \theta, \vartheta_a,
    \vartheta_b^{(1)}, \vartheta_b^{(2)}\Bigr)
.
\]
One would have equality here if the two datasets
$z_b^{(1)}$ and $z_b^{(2)}$ are independent,
which usually requires that the spatial domains in which
their respective forward processes take place are disjoint
(and far apart).
In practical applications, however, this is unlikely to be the case.
Note, in particular, that the dependence between $z_b^{(1)}$ and
$z_b^{(2)}$
is not diminished even if the two forward processes are of
incomparable physical natures, for example one is a pumping test
and the other is a tracer test.
The dependence is due to the common $Y$ field that underlies both tests.
Another situation in which one would have equality in the relation above is
that
$z_b^{(1)}$ is implied by
$\bigl(\theta, \vartheta_a, \vartheta_b^{(1)}, \vartheta_b^{(2)}\bigl)$,
that is,
any field $\tilde{y}$ conditional on any particular
$\bigl(\theta, \vartheta_a, \vartheta_b^{(1)}, \vartheta_b^{(2)}\bigr)$
would reproduce (i.e.\@ predict by the forward model)
$z_b^{(1)}$ exactly.
This can not happen in the current context.

Formula~(17) in \citet{Rubin:2010:BAI}
assumes $z_b^{(1)}$ and $z_b^{(2)}$ are independent.
As mentioned above,
this assumption could be valid when the respective forward processes for
$z_b^{(1)}$ and $z_b^{(2)}$ take place in disjoint and far-apart regions.
The further condition of ``far-apart'' is needed because
of spatial correlation in the $Y$ field.
When these conditions are met,
$z_b^{(1)}$ and $z_b^{(2)}$ would be disjointly useful for their
respective spatial regions.
In such situations,
it may not be appropriate to assume
the different regions share the same value of the structural parameter
$\theta$.
Consequently, one may as well need to divide the entire spatial domain
into isolated sub-regions and study them separately.
This is not a typical scenario of experimental or observational studies.

\citet[sec.~4.2]{Murakami:2010:DBG} makes similar attempts and has similar
problems.
More discussion on this issue can be found
in \citet[sec.~9]{Zhang:2008:DAS}.

\section{Comparison with the pilot point method}

Very early development of the anchor idea
benefited from the popular ``Pilot Point'' method (PPM)
\citep{deMarsily:1984:IIT}.
This is shown in that
\citet{Zhang:2008:DAS} used the name ``pilot point'' provisionally
in describing the new concept,
before the name ``anchor'' occurred to the author.
(The name ``pivot point'' occurred before ``anchor'' but was abandoned;
see \citet{Zhang:2008:APP1}.)
Some superficial similarities notwithstanding,
the two methods are very different;
see \citet{Zhang:2008:APP1} for some discussion.

As is explained in \citet[sec.~8.3]{Rubin:2010:BAI},
an important difference between PPM and anchored inversion is that
the former produces an ensemble of pilot point values that are not a
random sample of a clearly-defined target distribution,
whereas the latter infers the distribution of anchor values
(along with other parameters) following statistical rules.
These observations are borrowed from
\citet{Zhang:2008:APP1},
\citet{Zhang:2008:PPM},
\citet{Zhang:2008:APP2},
and
Zhang [personal correspondence to Rubin, 8 May 2008].
(See, e.g.,
paragraph~[104] of \citet{Rubin:2010:BAI} and
the 2nd paragraph of sec.~1 in \citet{Zhang:2008:APP2}.)

Another issue with PPM is that it requires the geostatistical parameters
($\theta$) to be pre-specified, and has no provisions for systematically
updating these parameters in light of the type~B data $z_b$
(\citet[page~5]{Zhang:2008:APP1} and \citet[sec.6.1]{Zhang:2009:MGR}).

In PPM,
the same optimization procedure is applied to
each (vector) value ever tried for the pilot points,
and each value for the pilot points eventually leads to
one realization of the field $Y$.
As a result,
the computational cost is proportional to the number of field realizations
created in the resultant ensemble.
In anchored inversion,
inference of the parameter (both $\theta$ and $\vartheta$) distribution
requires a large number of field realizations.
Once this inference is done,
creating field realizations (to be used for characterization, analysis,
prediction, etc.) is not part of the inversion procedure,
and the computational expense is essentially negligible.
This aspect is discussed in
\citet[secs.~6.1,~7.1]{Zhang:2009:MGR}

\section{Additional issues}

This section lists some technical inaccuracies
in \citet{Rubin:2010:BAI}.



In paragraph~[26] the authors state that
they ``could use
$p(\theta \given \vartheta_a)
\propto p(\vartheta_a \given \theta) p(\theta)$
to accommodate the prior'' of $\theta$.
This would be correct if and only if
$p(\vartheta_a)$ is constant, which is unlikely the case in this
context.

In paragraph~[32], the authors state that ``the prior for the entire model
appears indirectly in equation~(3) through the relation
$p(\theta \given \vartheta_a)
\propto p(\theta) p(\vartheta_a \given \theta)$.''
First,
as mentioned above,
this relation does not hold unless $p(\vartheta_a)$ is constant.
Second,
the model parameter vector contains $\theta$ and $\vartheta$,
hence a prior for the ``entire'' model is a function
of both $\theta$ and $\vartheta$.
In fact, the prior for the model is provided by
$p(\theta, \vartheta \given z_a)
= p(\vartheta_a \given z_a) p(\theta \given \vartheta_a, z_a)
  p(\vartheta_b \given \theta, \vartheta_a, z_a)$.
(It can be simplified if $z_a$ is error-free.)
This prior specification makes use of part of the data ($z_a$)
and leaves the other part ($z_b$) as the ``main'' data.
See
\citet[paragraphs~1--2 of sec.~6.1]{Zhang:2008:IMS1} or
\citet[paragraphs~{[21--22]}]{Rubin:2010:BAI}.


In paragraph~[37] appears this statement:
``When multiple data types are involved, that would include conditional
simulation of the $Y$ field followed by forward modeling.''
This could confuse the reader.
``Conditional simulation'' and ``forward modeling'' are the basic
components of the method; they need to be run whether or not
``multiple'' data ``types'' are involved.

Paragraph~[44] cites several reasons for the
discrepancy between the prediction $M(\tilde{y})$ and the observation $z_b$.
The cited reasons are not the most important ones.
The most important reason for this discrepancy is
that the field $Y$ conditional on fixed $\theta$ and $\vartheta$ is
random.
In fact, the method allows all four types of the cited ``errors'
to be absent, and in that case $M(\tilde{y})$ would still deviate
from $z_b$.

The confusing notation ``parametric error'' appears in
paragraph~[44] and the confusion is reinforced in paragraph~[45]
by the statement
``The parameter error is represented by $\epsilon_b$,
and it covers errors in $\vartheta_b$ and in the structural parameters
$\theta$''.
One knows from formula~(2) that $\epsilon_b$
indicates error in the type~B data.
This does not include ``errors'' in $\vartheta_b$ and
$\theta$.
The same misstatement appears in the ``Notation''.
Recall that $\vartheta_b$ and $\theta$ are model parameters in a Bayesian
approach, implying that the notion of their ``errors'' is undefined.

%
%

In paragraphs [68]--[71],
the authors replace the more conventional notation $\vec{\beta}$
(for linear coefficients) in \citet[Appendix]{Zhang:2008:IMS1}
by $\vec{m}$ and call it ``design matrix'',
causing profound confusion.
``Design matrix'' in statistics involves $x$ (spatial coordinates) only;
it is not an unknown parameter subject to estimation.
A better notation would be $\vec{\beta}$,
which multiplies the ``design matrix''
($\mat{X}$ in \citet[Appendix]{Zhang:2008:IMS1})
to get the mean of the spatial variable.
When the field has a constant mean,
the design matrix would be a column matrix of 1's
and $\vec{\beta}$ would be of length 1.
This is not like what is stated in paragraph~[69],
``all the terms in $\vec{m}$ are equal to $\mu$''.
The $\vec{m}$ in formula~(12) is meant to be the mean of $Y$ at locations
$\xa$.
This mean is a function of the structural parameters
(in particular, the trend parameters)
as well as the location vector $\xa$.
It is a vector of length $n_a$, and is not
the model parameter $\vec{m}$ in the previous paragraphs.
(The length of the model parameter vector is independent of $n_a$.)
However, its functional relation with the structural parameters
is lost in~(12).
A similar confusion occurs in formula~(14).

Following the choice of symbols in ``Notation'',
all the $\xb$ in section~6.1.2 should be
$\vec{x}_{\vartheta_b}$, that is, it is the locations of the anchors
$\vartheta_b$.
This is not $\xb$, which is the locations of type~B data $\zb$
(as mentioned above, however, the ``location of type~B data'' is not
always meaningful in this method).
The $\xa$ should be $\vec{x}_{\vartheta_a}$,
namely the locations of the anchors $\vartheta_a$;
but these locations coincide with $\xa$.

In formula~(13),
the $\za$ between the pair of `$\lVert$'s should be $\vartheta_b$.
The formula as is written is not a function of $\vartheta_b$.

In formula~(14), the $\zb$ should be $\za$.
The former is not available in this context and is not related to the
quantities the way stated in~(14).
The $\vec{m}$ is a function of $\theta$ and $\vec{x}_{\vartheta_b}$,
giving the mean of $Y$ at locations $\vec{x}_{\vartheta_b}$.
The confusion about $\vec{m}$ has been mentioned earlier.

In formula~(14),
the $\mat{R}_{\xb\given \xa}$
should be $\mat{R}_{\vec{x}_{\vartheta_b}, \xa}$.
This is the correlation between $\vartheta_b$ and $\vartheta_a$,
not the ``conditional'' correlation of $\vartheta_b$ given
$\vartheta_a$,
which is the $\mat{R}_{\xb \given \xa}$ in
(13) and~(15).

The $\mat{R}_{x_b \given x_a}$ given
in~(15) is the conditional ``correlation'' of $\vartheta_b$,
not conditional ``covariance'' as claimed before formula~(14).
In \citet[Appendix]{Zhang:2008:IMS1},
the counterpart to~(15) is written for the conditional covariance,
$\sigma^2 \mat{R}$.

The formulas~(11)--(15) are taken from the Appendix of
\citet{Zhang:2008:IMS1}, with necessary adjustment because
the Box-Cox transform used in
\citet{Zhang:2008:IMS1} is dropped in
\citet{Rubin:2010:BAI}.
The two papers also differ slightly in choice of symbols.
It is not possible to give correct versions of
(12), (14), and (15) without introducing symbols that are not defined
in sections 6.1.1 and~6.1.2,
because the concept of ``design matrix'' and the symbol $\vec{m}$
defined in paragraph~[68] are problematic.

The right-hand side of formula~(18) equals
$p(\za, \zb \given \theta)$, which is the likelihood.
Optimizing the likelihood would lead to ML rather than MAP estimates.
Formula~(18) would be correct if the prior $p(\theta)$ were flat.
However, that would make the formulation non-Bayesian,
and hence unrelated to MAP.
In paragraph~[90],
$p(\za \given \theta)$ is referred to as ``prior''.
However, $\za$ are data, which do not have ``prior''.
In paragraph~[91] appears the relation
$\epsilon_b = \zb - \mathcal{M}(\tilde{y})$.
However, $\tilde{y}$ does not appear in~(18),
hence $\epsilon_b$ is undefined.

\section{Conclusion}

The method of anchored inversion has a number of notable features
including
(1) the anchor concept,
which on the one hand creates a bridge between the spatial attribute field $Y$
and the forward data $\zb$,
and on the other hand separates the dimensionality of the model parameter
($\theta$ and $\vartheta$) from that of the numerical grid;
and
(2) the recognition that the likelihood function is unknown
and needs to be estimated nonparameterically.
The method is general and in a sense modular,
although these desirable features are not unique to this method.

The anchor parameterization may be used in a Bayesian way
(e.g.\@ deriving posterior distributions) or a non-Bayesian way
(e.g.\@ obtaining maximum likelihood estimates).
\citet{Zhang:2008:IMS1} presented a Bayesian procedure for using anchors.

Additional comments on this method can be found in
\citet{Zhang:2010:ICB}.

To conclude with a little history,
the method (``anchored inversion'') as introduced in
\citet{Zhang:2008:DAS} was fully described and implemented in
\citet{Zhang:2008:IMS1}.
The same method was re-presented (by copy-pasting and rephrasing) in
\citet{Rubin:2010:BAI}
without the knowledge and consent of Zhang, the actual author,
but with technical errors and misinterpretations.
The method was fundamentally changed in
\citet{Zhang:2009:MGR}, which has turned out to be an intermediate
milestone,
as the method(s) described in these works has now been superseded
by a new, and superior, method described in \citet{Zhang:2011:AAI}.

\section{Postscript}

The line of scandalous practice in
\citet{Rubin:2010:BAI}
has been continued by
\citet{Murakami:2010:DBG},
\citet{Murakami:2010:BAT},
\citet{Chen:2012:TDB},
\citet{Osorio-Murillo:2012:EIM},
\citet{Yang:2012:SPL},
\citet{Over:2013:SIC},
\citet{Osorio-Murillo:2014:DII},
\citet{Hebe:2015:CIR},
\citet{Osorio-Murillo:2015:SFI},
and
\citet{Over:2015:BIM}.

With the unfortunate prospect of propagating
misinterpretation and misattribution,
\citet{Rubin:2010:BAI} and its follow-ups
have been referenced in a number of publications,
including
\citet{%
Bellin:2016:CCH,
Castagna:2015:UEE,
Chen:2012:SME,
Chen:2013:AEB,
Crestani:2013:EKF,
Cui:2014:CPS,
Dafflon:2011:HPE,
Dai:2016:DWA,
deBarros:2012:IHD,
deBarros:2013:CSM,
FernandezGarcia:2012:BAI,
Gao:2016:IRD,
Hesse:2014:JIC,
Kitanidis:2014:PCG,
Kitanidis:2015:PQH,
Koch:2014:CCS,
Kowalsky:2012:PIP,
Laloy:2015:PIM,
Leube:2012:BAE,
Liu:2016:EML,
Lu:2012:MBA,
Lu:2013:EEC,
Nowak:2012:HDA,
Piccolroaz:2015:OUS,
Pool:2015:CDS,
Rogiers:2013:UOA,
Shi:2014:APU,
Williams:2011:PSU,
Xue:2014:MDA,
Yoon:2013:HPI,
Zachara:2013:RTU,
Zhang:2013:ASG,
Zhou:2012:PSB,
Zhou:2014:IMH}.

\end{document}